\PassOptionsToPackage{utf8}{inputenc}
\documentclass[]{article}

\usepackage{graphicx}
\usepackage{url}
\begin{document}
\title{Design and Development of PCN-Miner: A tool for the Analysis of Protein Contact Networks.}

\maketitle

\abstract{\textbf{Motivation:}
Protein Contact Network (PCN) is a powerful tool for analysing the structure and function of proteins. In particular, PCN has been used for disclosing the molecular features of allosteric regulation through PCN clustering. Such analysis is  relevant in many applications, such as the recent study of SARS-CoV-2 Spike Protein. Despite its relevance, methods for the analysis of PCN are  spread into a set of different libraries and tools. Therefore, the introduction of a tool that incorporates all the function may help researchers.\\ 
\textbf{Results:} We present PCN-Miner a software tool implemented in  the Python programming language able to import protein in the Protein Data Bank format and generate the corresponding protein contact network. Then it offers a set of algorithms for the analysis of PCS that cover a large set of applications: from clustering to embedding and subsequent analysis. \\
\textbf{Availability:} Software is available at \url{https://github.com/hguzzi/ProteinContactNetworks} \\
\textbf{Contact:} Pietro Hiram Guzzi \\
\textbf{Supplementary information:} Supplementary data are available at \url{https://github.com/hguzzi/ProteinContactNetworks}}
\section{Introduction}
Proteins exert a central role in biology through a unique, tight relationship between their molecular structure and function.
Structural information about protein molecules are derived both from experiments (X-ray crystallography or NMR, among others \cite{biswas2013proteins}) and from computational methods \cite{petrey2005protein}. The freely accessible Protein Data Bank gathers protein structural information in a specific format (PDB file); the plenty of structural data information requires prediction and analysis tools so to unravel the structure-function relationship, to identify emerging features and to predict biological mechanisms involving protein molecules \cite{eswar2003tools}.

In this framework, Protein Contact Networks (PCNs) emerged as a relevant paradigm for the analysis of protein molecular structures \cite{di2013protein}. 

PCN descriptors are useful to frame protein function, with a special regard to properties linked to protein modularity, such as allosteric regulation, e.g. One of the key feature PCNs are able to catch is the protein structure modularity at the basis of domain functional partition of protein molecular structures and allosteric regulation.  \cite{khan2015modularity}. PCNs model allows to identify modules in protein molecules through network spectral clustering \cite{tasdighian2014modules, Di-Paola:2015un}, with relevant application in different biological contexts (\cite{Cimini:2016wl, Di-Paola:2020ww}.

PCNs are built up starting from the structural information in the PDB files, extracting the position of the residues alpha-carbons and computing the distance matrix, whose generic element $d_{ij}$ is the Euclidean distance between the alpha-carbons of the i-th and j-th residues. Protein network nodes are the single residues and link exists between nodes (residues) if their distance lies between 4 and 8 \AA, to include only relevant noncovalent intramolecular interactions. Thus, the generic element of the PCN adjacency matrix $A_{ij}$ is 1 if a link exists between the i-th and the j-the residues, 0 otherwise \cite{di2021disclosing}. 

\begin{figure}[ht]
\centering
\includegraphics[width=0.9\columnwidth]{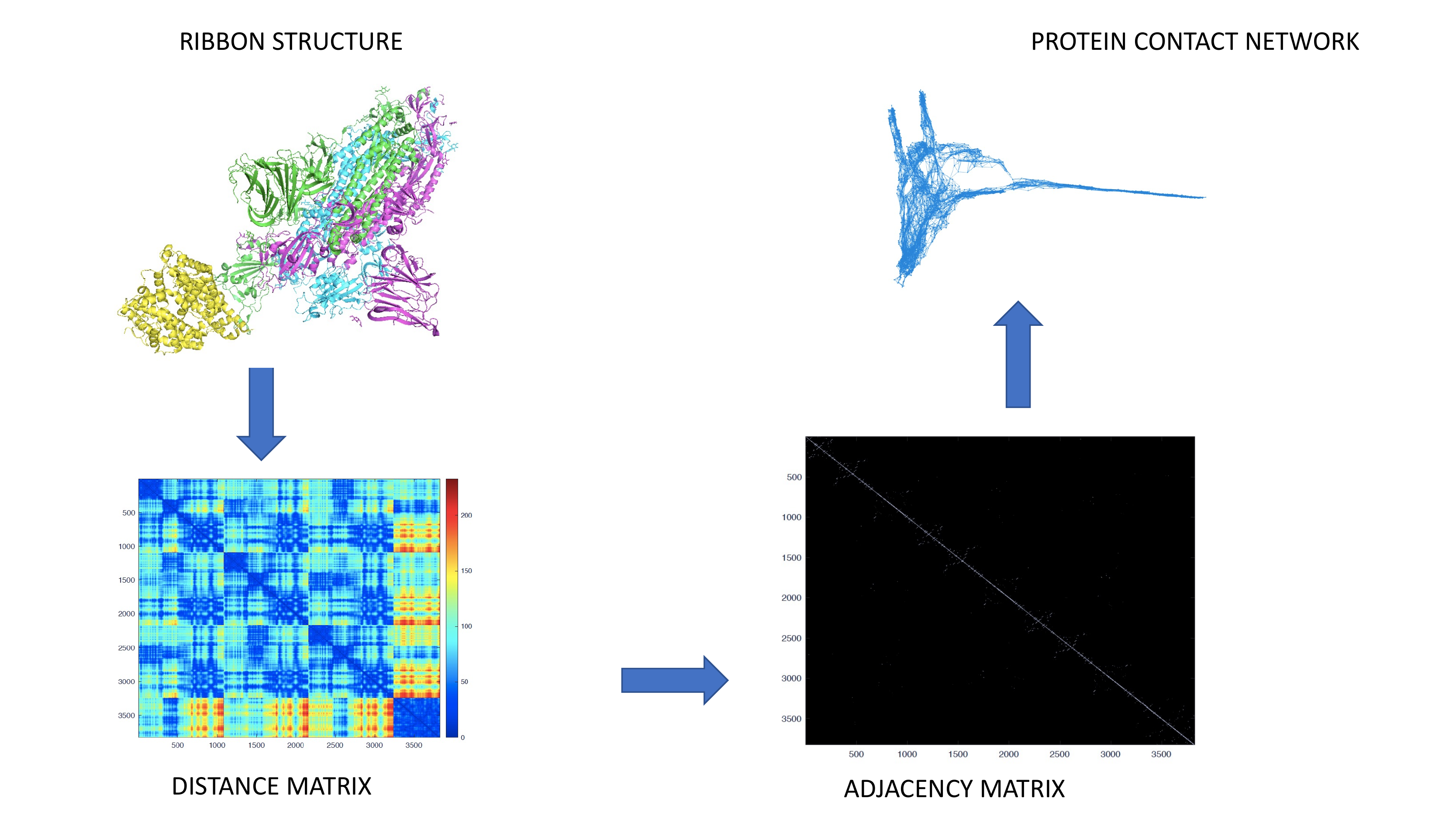}
\label{fig:PCN}
\caption{Scheme of the PCN construction on the close conformation of SARS-CoV 2 spike protein (PDB code 6vyb): starting from the structure (upper left) it is possible to compute the distance matrix (lower left) and so the adjacency matrix (lower right) so finally the PCN (upper right)}
\end{figure}

Figure \ref{fig:PCN} reports the logical scheme of the PCN construction shown with reference to the close conformation of the SARS-CoV2 spike protein (PDB code 6vyb).


Starting from adjacency matrices, it is possible to extract biologically relevant knowledge using algorithms coming from graph theory. 

For instance, Allostery is the ability of proteins to transduce a signal from a from a sensor (allosteric) site to the effector (binding) site, even if they are not in direct contact. In this way, proteins can perceive ligand binding at distance
From the active site, or to perceive other perturbations or external stimuli  \cite{di2020discovery}.

It has been shown that PCNs offer a theoretical framework to study allostery since they can encode the mechanism of transmission of signal. In particular, in \cite{di2020discovery} author shows the analysis of network modules through spectral clustering \cite{von2007tutorial} can reveal allostery. The study of allostery has recently shown a big impact on the study of conformational structure and of the binding related to SARS-CoV-2 Spike proteins \cite{dubanevics2021computational,ortuso2021structural,guzzi2020master}.

Unfortunately, there is a lack of tools that enable an easy study of PCNs. To the best of our knowledge, existing libraries for the whole workflow of the analysis are spread in a set of different libraries able to perform the main step of the analysis: (i) reading protein structures encoded into database files, (ii) building protein contact networks, (iii) analysis of PCN and (iv) visualisation of the obtained results \cite{di2013protein}.
For these aims, we designed and implemented PCN-Miner a software tool able to implement these functions in a single package. PCN-Miner offer main functionalities for reading and analysing PCNs. 
Contributions of the software are:
(i) to identify the putative allosteric paths and regions in protein structures so being of help in the design of allosteric drugs; (ii) to allow for hypothesis generation on the functional effect of mutations; and (iii) to recognise funtional domains in proteins.

\section{Methods}
PCN-Miner is implemented in Python 3.8 programming language. It uses  scipy and numpy libraries for managing matrices. Management of PDB files is provided by ProDy package. The network embedding is realised by wrapping the GEM library and clustering algorithms by CdLib (see supplementary materials for all details about libraries). Visualisation of protein structures is made by wrapping the community edition of PyMol. 
The current version of PCN-Miner wraps the NetworkX,  cdlib library and GEM libraries \cite{goyal2018graph}.

\section{Results}
\begin{figure}[ht]
\centering
\includegraphics[width = 3in,height=3.3in]{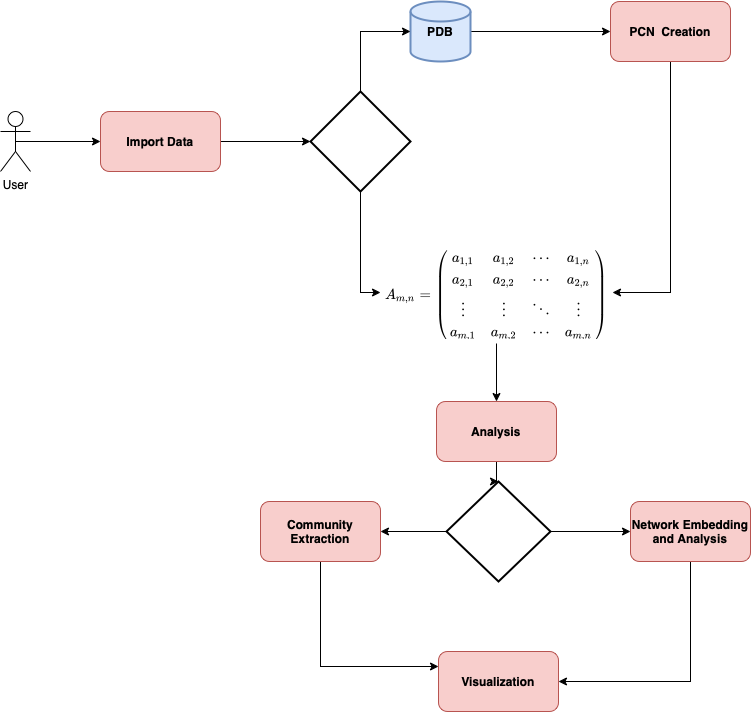}
\label{fig:workflow}
\caption{Workflow of analysis of PCN in PCN-Miner. The user can import PDB files and then he/she can obtain the PCN or alternatively he/she can directly import a PCN previously determined. Then he can access the analysis functionalities (e.g. community extraction on PCN or embedding), then he can visualise the obtained results on the protein.}
\end{figure}

As a results user may easily implement the whole workflow of analysis as depicted in Figure \ref{fig:workflow}. User can import PDB files and then he/she can obtain the PCN or alternatively he/she can directly import a PCN previously determined. Then he can access the analysis functionalities (e.g. clustering, community extraction on PCN or embedding).
Current version of PCN-Miner implements both soft and hard clustering of adjacency matrices . Moreover, it implements both hard and soft clustering  on un-normalised, normalised an Shi-Malik normalised laplacian matrices. In parallel the user may analyse communities on PCN by calling one of the following algorithms: 
Louvain, Leiden, Spinglass and Walktrap, asynchronous fluid community algorithm (asyn\_fluid), and  the Clauset-Newman-Moore greedy modularity maximization (greedy\_modularity) .


It is also able to map PCN into an embedding subspace through  HOPE \cite{ou2016asymmetric} and LaplacianEigenmaps embedding algorithms. After that embedding is done, the user may use clustering algorithms to analyse such space \cite{belkin2003laplacian}.

%
%
\section{Conclusion}
We presented  a tool written in Python  to import protein in Protein Data Bank format and generate corresponding protein contact network. We showed the ability of our software to implement a whole workflow of analysis from clustering to embedding and subsequent analysis.
\section*{Acknowledgements}
Authors thank Eng. Ugo Lomoio and Eng. Giuseppe Ferrarelli for their work on developing software modules.

\section{Authors}

Pietro Hiram Guzzi and Pierangelo Veltri are within Department of Surgical and Medical Sciences, University of Catanzaro.

Luisa Di Paola is Unit of Chemical-Physics Fundamentals in Chemical Engineering, Department of Engineering, University Campus Bio-Medico di Roma, via Ãlvaro del Portillo 21, 00128 Rome, Italy.

Alessandro Giuliani is within Environment and Health Department, Istituto Superiore di Sanità.

\section*{Funding}

PHG and PV were partially funded by PON-VQA project.

\bibliographystyle{unsrt}
\bibliography{document.bib}
\end{document}